\documentclass{siamonline0516}

\usepackage{lipsum}
\usepackage{amsfonts}
\usepackage{tikz}
\usepackage{graphicx}
\usepackage{epstopdf}
\usepackage{algorithmic}
\usepackage{bbm}
\ifpdf
  \DeclareGraphicsExtensions{.eps,.pdf,.png,.jpg}
\else
  \DeclareGraphicsExtensions{.eps}
\fi

\newcommand{\TheTitle}{Spontaneous wave formation in stochastic self-driven particle systems} 
\newcommand{\TheAuthors}{M. Friesen, H. Gottschalk, B. R\"udiger, A. Tordeux}

\headers{Spontaneous wave formation in stochastic self-driven particle systems}{\TheAuthors}

\title{{\TheTitle}
}

\author{
  Martin Friesen
		\and
  Hanno Gottschalk
  \and
  Barbara R\"udiger\thanks{School of Mathematics and Natural Sciences, University of Wuppertal, Germany
    (\email{friesen@math.uni-wuppertal.de}, \email{hanno.gottschalk@uni-wuppertal.de},  \email{ruediger@uni-wuppertal.de}).}
    \and
    Antoine Tordeux\thanks{School of Mechanical Engineering and Safety Engineering, University of Wuppertal, Germany
    (\email{tordeux@uni-wuppertal.de}).}
}

\usepackage{amsopn}

\DeclareMathOperator*{\fle}{\rightarrow}

\DeclareMathOperator*{\R}{\mathbb{R}}

\newcommand{\dd}{\,d}
\newcommand{\tra}{^\top}

\newcommand{\ba}{\begin{array}{ll}}
\newcommand{\ea}{\end{array}}

\newcommand{\be}{\begin{equation}}
\newcommand{\ee}{\end{equation}}

\newcommand{\bi}{\begin{itemize}}
\newcommand{\ei}{\end{itemize}}

\newcommand{\bc}{\begin{center}}
\newcommand{\ec}{\end{center}}

\newcommand{\bfig}{\begin{figure}[!ht]}
\newcommand{\efig}{\end{figure}}

\newcommand{\ben}{\begin{enumerate}}
\newcommand{\een}{\end{enumerate}}

\newcommand{\bmat}{\left[\begin{matrix}}
\newcommand{\emat}{\end{matrix}\right]}

\newcommand{\1}{\mathbbm{1}}

\ifpdf
\hypersetup{
  pdftitle={\TheTitle},
  pdfauthor={\TheAuthors}
}
\fi

\begin{document}

\maketitle

\begin{abstract}
Waves and oscillations are commonly observed in the dynamics of self-driven agents such as pedestrians or vehicles.
Interestingly, many factors may perturb the stability of space homogeneous streaming, leading to the spontaneous formation of collective oscillations of the agents related to stop-and-go waves, jamiton, or phantom jam in the literature. 
In this article, we demonstrate that even a minimal additive stochastic noise in stable first-order dynamics can initiate stop-and-go phenomena. 
The noise is not a classic white one, but a colored noise described by a Gaussian Ornstein-Uhlenbeck process. 
It turns out that the joint dynamics of particles and noises forms again a (Gaussian) Ornstein-Uhlenbeck process whose characteristics can be explicitly expressed in terms of parameters of the model.
We analyze its stability and characterize the presence of waves through oscillation patterns in the correlation and autocorrelation of the distance spacing between the particles. 
We determine exact solutions for the correlation functions for the finite system with periodic boundaries and in the continuum limit when the system size is infinite.
Finally, we compare experimental trajectories of single-file pedestrian motions to simulation results.
\end{abstract}


\tableofcontents
\bigskip\medskip

\section{Introduction}
The emergence of collective motion behaviors is frequently observed in the dynamics of agents interacting locally. 
Examples are swarming and the formation of patterns and structures in bacterial colonies, animal aggregations, or traffic flow and pedestrian dynamics \cite{Buhl2006,Ben-Jacob1994,Vicsek2012,Helbing1997,Helbing2000b}. 
Spontaneous formation of stop-and-go waves in uni-directional road traffic or pedestrian streams is a typical example of self-organization. 
Stop-and-go phenomena, also related to accordion-like traffic, phantom jam, jamiton, or self-sustained waves in the literature \cite{Kurtze1995,Seibold2013,Flynn2009}, currently occur in vehicle, pedestrian or again bicycle flows \cite{Orosz2010,Boltes2018}. 
The flows in congested states tend to stream jerky with acceleration and deceleration phases instead of streaming uniformly. 
Stop-and-go waves even emerge in single-file experiments where neither the infrastructure nor the initial configuration can explain their presence \cite{Sugiyama2008,Zhang2014,STERN2018}. 
Beside scientific interests, stop-and-go waves impact the safety and the comfort of the users, and also the environment. 
Indeed, they generate more fuel consumption and pollutant emission than space homogeneous streaming \cite{Aguilera2014,Stern2019}. 

Road traffic and pedestrian flow models are microscopic, mesoscopic or macroscopic. 
Microscopic approaches describe individual trajectories with following models and agent-based approaches. 
Mesoscopic models are gas-kinetic frameworks describing probability density functions for the speeds and agent positions, while macroscopic models are partial differential equations for aggregated performances (see \cite{Chowdhury2000,Bellomo2011,van2015} for reviews). 
The well-known Lighthill-Whitham-Richards macroscopic model \cite{Richards1956,Lighthill1955} describes for Riemann problems shock and rarefaction waves propagating at speeds given by the Rankine-Hugoniot formula. 
Yet, the model is first-order and it fails to explain the auto-organisation in waves of perturbed systems.
Generally speaking, the spontaneous formation of stop-and-go waves requires inertial second order frameworks and the use of delayed processes, see, the references \cite{Bando1995,Barlovic1998,Jiang2001,Davis2003} 
for microscopic models, \cite{Helbing1998,Bellomo2012}  
for mesoscopic models, or \cite{Colombo2003,Goatin2006,Seibold2013} 
for macroscopic ones. 
The emergence of stop-and-go waves is explained through instability of space homogeneous solutions, the stability breaking down when delay or relaxation times (i.e.\ inertia) exceed critical thresholds \cite{Orosz2009,Orosz2010}. 
In the unstable case, the solutions can be periodic, quasi-periodic, limit cycle or even chaotic dynamics with stop-and-go waves \cite{Tomer2000}. 
Derivations in macroscopic hyperbolic continuum are Korteweg-de Vries, modified Korteweg-de Vries or time-dependent Ginzburg-Landau soliton equations \cite{Muramatsu1999,Nagatani1998,Bellomo2008,Aw2002}. 

In this article, we demonstrate that stop-and-go waves even emerge from stochastic noise effects without requiring instability phenomena. 
Generally speaking, the introduction of white noises tends to increase disorder and prevent self-organization
\cite{Vicsek1995,Helbing2000}, while coloured noises can generate complex structures and patterns \cite{Arnold1978,Castro1995}. 
In most of self-driven agent models, the noises added to the dynamics are white \cite{Helbing1995,Tomer2000,Buhl2006,Helbing2000}. 
We show in this article that the introduction of a particular colored noise in stable first-order dynamics can initiate collective oscillations in the system and spontaneous formation of stop-and-go waves. 
The noise is generated by a Gaussian Ornstein-Uhlenbeck process. 
The choice of such a colored noise is motivated by statistical evidence showing linear shapes of the spectral density of pedestrian speed in square inverse frequency domain \cite{Tordeux2020}.
The waves are characterised by analysing the correlation and autocorrelation functions of the particle spacing describing characteristic oscillating patterns \cite{Bain2019}.
In contrast to classical inertial deterministic approaches, neither instability nor phase transition phenomena are observed.
This makes the stochastic approach more convenient to analyse. 
Indeed, the system is Gaussian and ergodic, i.e.\ admitting a unique invariant measure for any initial condition.

The stochastic model has been introduced to describe by simulation stop-and-go waves in pedestrian dynamics \cite{Tordeux2016}. 
We propose in this article to rigorously demonstrate the presence of waves by analysing the structure of the correlation and autocorrelation functions and their periodic characteristics. 
We carry out the analysis for a finite system with periodic boundary conditions and at the limit of an infinite system.
The article is organised as following.
The stochastic model is defined in the next section. 
We solve the model in Sec.~\ref{solv} and analyse its stability in Sec.~\ref{stab}. 
The covariance functions are determined for a finite system with periodic boundaries in Sec.~\ref{covN}, and at the limit of an infinite system in Sec.~\ref{cova}. 
Finally, we compare simulation results to experimental data of pedestrian single-file motions in Sec.~\ref{sim}.

\section{Stochastic following model}\label{def}
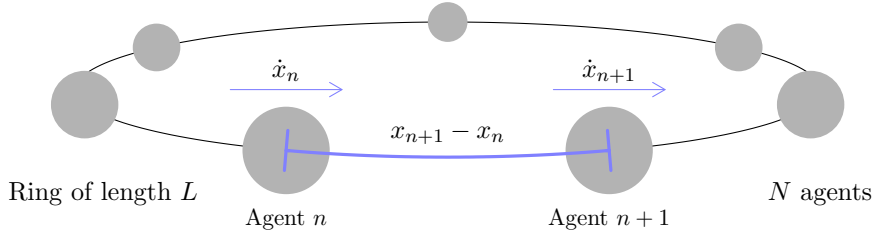
\begin{figure}[!ht]
\begin{center}\vspace{-4mm}
\begin{tikzpicture}[x=1.5pt,y=1.5pt]
\definecolor{fillColor}{RGB}{255,255,255}
\path[use as bounding box,fill=fillColor,fill opacity=0.00] (0,0) rectangle (247.08, 77.21);
\begin{scope}
\path[clip] ( 5.50, 5.50) rectangle (235.08, 65.21);
\definecolor{drawColor}{RGB}{0,0,0}

\path[draw=drawColor,line width= 0.4pt,line join=round,line cap=round] (123.54, 58.14) --
	(126.50, 58.14) --
	(129.46, 58.11) --
	(132.42, 58.07) --
	(135.36, 58.01) --
	(138.30, 57.93) --
	(141.22, 57.84) --
	(144.12, 57.73) --
	(147.00, 57.60) --
	(149.86, 57.46) --
	(152.69, 57.30) --
	(155.50, 57.13) --
	(158.27, 56.94) --
	(161.00, 56.73) --
	(163.70, 56.51) --
	(166.36, 56.27) --
	(168.98, 56.02) --
	(171.55, 55.75) --
	(174.07, 55.47) --
	(176.54, 55.18) --
	(178.96, 54.87) --
	(181.33, 54.54) --
	(183.64, 54.21) --
	(185.88, 53.86) --
	(188.07, 53.49) --
	(190.19, 53.12) --
	(192.24, 52.73) --
	(194.23, 52.34) --
	(196.14, 51.93) --
	(197.99, 51.51) --
	(199.76, 51.08) --
	(201.45, 50.63) --
	(203.06, 50.19) --
	(204.60, 49.73) --
	(206.05, 49.26) --
	(207.43, 48.78) --
	(208.72, 48.30) --
	(209.92, 47.81) --
	(211.04, 47.31) --
	(212.07, 46.81) --
	(213.01, 46.30) --
	(213.87, 45.79) --
	(214.63, 45.27) --
	(215.30, 44.75) --
	(215.88, 44.22) --
	(216.37, 43.69) --
	(216.77, 43.16) --
	(217.07, 42.63) --
	(217.28, 42.09) --
	(217.40, 41.56) --
	(217.42, 41.02) --
	(217.35, 40.48) --
	(217.19, 39.95) --
	(216.93, 39.41) --
	(216.58, 38.88) --
	(216.14, 38.35) --
	(215.60, 37.82) --
	(214.98, 37.30) --
	(214.26, 36.78) --
	(213.45, 36.26) --
	(212.55, 35.75) --
	(211.57, 35.25) --
	(210.49, 34.75) --
	(209.33, 34.25) --
	(208.08, 33.77) --
	(206.75, 33.29) --
	(205.34, 32.81) --
	(203.84, 32.35) --
	(202.27, 31.90) --
	(200.61, 31.45) --
	(198.88, 31.02) --
	(197.07, 30.59) --
	(195.20, 30.18) --
	(193.24, 29.77) --
	(191.22, 29.38) --
	(189.14, 29.00) --
	(186.98, 28.63) --
	(184.77, 28.27) --
	(182.49, 27.93) --
	(180.15, 27.60) --
	(177.76, 27.28) --
	(175.31, 26.98) --
	(172.82, 26.69) --
	(170.27, 26.42) --
	(167.67, 26.16) --
	(165.04, 25.91) --
	(162.36, 25.68) --
	(159.64, 25.47) --
	(156.88, 25.27) --
	(154.10, 25.09) --
	(151.28, 24.92) --
	(148.43, 24.77) --
	(145.56, 24.64) --
	(142.67, 24.52) --
	(139.76, 24.42) --
	(136.83, 24.34) --
	(133.89, 24.27) --
	(130.94, 24.22) --
	(127.98, 24.18) --
	(125.02, 24.17) --
	(122.06, 24.17) --
	(119.09, 24.18) --
	(116.14, 24.22) --
	(113.18, 24.27) --
	(110.24, 24.34) --
	(107.32, 24.42) --
	(104.40, 24.52) --
	(101.51, 24.64) --
	( 98.64, 24.77) --
	( 95.80, 24.92) --
	( 92.98, 25.09) --
	( 90.19, 25.27) --
	( 87.44, 25.47) --
	( 84.72, 25.68) --
	( 82.04, 25.91) --
	( 79.40, 26.16) --
	( 76.81, 26.42) --
	( 74.26, 26.69) --
	( 71.76, 26.98) --
	( 69.32, 27.28) --
	( 66.92, 27.60) --
	( 64.59, 27.93) --
	( 62.31, 28.27) --
	( 60.09, 28.63) --
	( 57.94, 29.00) --
	( 55.85, 29.38) --
	( 53.83, 29.77) --
	( 51.88, 30.18) --
	( 50.00, 30.59) --
	( 48.20, 31.02) --
	( 46.47, 31.45) --
	( 44.81, 31.90) --
	( 43.24, 32.35) --
	( 41.74, 32.81) --
	( 40.33, 33.29) --
	( 38.99, 33.77) --
	( 37.75, 34.25) --
	( 36.59, 34.75) --
	( 35.51, 35.25) --
	( 34.52, 35.75) --
	( 33.63, 36.26) --
	( 32.82, 36.78) --
	( 32.10, 37.30) --
	( 31.47, 37.82) --
	( 30.94, 38.35) --
	( 30.49, 38.88) --
	( 30.14, 39.41) --
	( 29.89, 39.95) --
	( 29.72, 40.48) --
	( 29.65, 41.02) --
	( 29.68, 41.56) --
	( 29.79, 42.09) --
	( 30.00, 42.63) --
	( 30.31, 43.16) --
	( 30.70, 43.69) --
	( 31.19, 44.22) --
	( 31.78, 44.75) --
	( 32.45, 45.27) --
	( 33.21, 45.79) --
	( 34.06, 46.30) --
	( 35.01, 46.81) --
	( 36.04, 47.31) --
	( 37.16, 47.81) --
	( 38.36, 48.30) --
	( 39.65, 48.78) --
	( 41.02, 49.26) --
	( 42.48, 49.73) --
	( 44.01, 50.19) --
	( 45.63, 50.63) --
	( 47.32, 51.08) --
	( 49.09, 51.51) --
	( 50.93, 51.93) --
	( 52.85, 52.34) --
	( 54.83, 52.73) --
	( 56.89, 53.12) --
	( 59.01, 53.49) --
	( 61.19, 53.86) --
	( 63.44, 54.21) --
	( 65.75, 54.54) --
	( 68.11, 54.87) --
	( 70.53, 55.18) --
	( 73.01, 55.47) --
	( 75.53, 55.75) --
	( 78.10, 56.02) --
	( 80.72, 56.27) --
	( 83.38, 56.51) --
	( 86.07, 56.73) --
	( 88.81, 56.94) --
	( 91.58, 57.13) --
	( 94.38, 57.30) --
	( 97.22, 57.46) --
	(100.07, 57.60) --
	(102.96, 57.73) --
	(105.86, 57.84) --
	(108.78, 57.93) --
	(111.71, 58.01) --
	(114.66, 58.07) --
	(117.61, 58.11) --
	(120.57, 58.14) --
	(123.54, 58.14);
\definecolor{fillColor}{gray}{0.70}

\path[fill=fillColor] ( 82.80, 25.85) circle ( 11);

\path[fill=fillColor] (164.27, 25.85) circle ( 11);

\path[fill=fillColor] (215.07, 37.37) circle ( 8.5);

\path[fill=fillColor] (196.94, 51.75) circle (  6);

\path[fill=fillColor] (123.54, 58.14) circle (  5);

\path[fill=fillColor] ( 50.13, 51.75) circle (  6);

\path[fill=fillColor] ( 32.00, 37.37) circle ( 8.5);
\definecolor{drawColor}{RGB}{128,128,255}

\path[draw=drawColor,line width= 1.2pt,line join=round,line cap=round] ( 82.80, 25.85) --
	( 83.57, 25.78) --
	( 84.34, 25.72) --
	( 85.12, 25.65) --
	( 85.90, 25.59) --
	( 86.68, 25.53) --
	( 87.46, 25.47) --
	( 88.25, 25.41) --
	( 89.04, 25.35) --
	( 89.83, 25.30) --
	( 90.63, 25.24) --
	( 91.43, 25.19) --
	( 92.23, 25.14) --
	( 93.03, 25.09) --
	( 93.84, 25.04) --
	( 94.65, 24.99) --
	( 95.46, 24.94) --
	( 96.27, 24.90) --
	( 97.09, 24.85) --
	( 97.91, 24.81) --
	( 98.73, 24.77) --
	( 99.55, 24.73) --
	(100.37, 24.69) --
	(101.20, 24.65) --
	(102.02, 24.62) --
	(102.85, 24.58) --
	(103.69, 24.55) --
	(104.52, 24.52) --
	(105.35, 24.49) --
	(106.19, 24.46) --
	(107.03, 24.43) --
	(107.86, 24.40) --
	(108.70, 24.38) --
	(109.55, 24.35) --
	(110.39, 24.33) --
	(111.23, 24.31) --
	(112.08, 24.29) --
	(112.92, 24.27) --
	(113.77, 24.26) --
	(114.61, 24.24) --
	(115.46, 24.23) --
	(116.31, 24.22) --
	(117.16, 24.20) --
	(118.01, 24.19) --
	(118.86, 24.19) --
	(119.71, 24.18) --
	(120.56, 24.17) --
	(121.41, 24.17) --
	(122.26, 24.17) --
	(123.11, 24.16) --
	(123.96, 24.16) --
	(124.82, 24.17) --
	(125.67, 24.17) --
	(126.52, 24.17) --
	(127.37, 24.18) --
	(128.22, 24.19) --
	(129.07, 24.19) --
	(129.92, 24.20) --
	(130.77, 24.22) --
	(131.62, 24.23) --
	(132.46, 24.24) --
	(133.31, 24.26) --
	(134.16, 24.27) --
	(135.00, 24.29) --
	(135.85, 24.31) --
	(136.69, 24.33) --
	(137.53, 24.35) --
	(138.37, 24.38) --
	(139.21, 24.40) --
	(140.05, 24.43) --
	(140.89, 24.46) --
	(141.72, 24.49) --
	(142.56, 24.52) --
	(143.39, 24.55) --
	(144.22, 24.58) --
	(145.05, 24.62) --
	(145.88, 24.65) --
	(146.71, 24.69) --
	(147.53, 24.73) --
	(148.35, 24.77) --
	(149.17, 24.81) --
	(149.99, 24.85) --
	(150.81, 24.90) --
	(151.62, 24.94) --
	(152.43, 24.99) --
	(153.24, 25.04) --
	(154.04, 25.09) --
	(154.85, 25.14) --
	(155.65, 25.19) --
	(156.45, 25.24) --
	(157.24, 25.30) --
	(158.04, 25.35) --
	(158.83, 25.41) --
	(159.62, 25.47) --
	(160.40, 25.53) --
	(161.18, 25.59) --
	(161.96, 25.65) --
	(162.73, 25.72) --
	(163.51, 25.78) --
	(164.27, 25.85);

\path[draw=drawColor,line width= 1.2pt,line join=round,line cap=round] ( 82.80, 25.85) -- ( 83.57, 25.78);

\path[draw=drawColor,line width= 1.2pt,line join=round,line cap=round] ( 83.24, 30.89) --
	( 82.80, 25.85) --
	( 82.37, 20.81);

\path[draw=drawColor,line width= 1.2pt,line join=round,line cap=round] (164.27, 25.85) -- (163.51, 25.78);

\path[draw=drawColor,line width= 1.2pt,line join=round,line cap=round] (164.71, 20.81) --
	(164.27, 25.85) --
	(163.84, 30.89);
\definecolor{drawColor}{RGB}{0,0,0}


\definecolor{drawColor}{RGB}{128,128,255}

\path[draw=drawColor,line width= 0.4pt,line join=round,line cap=round] ( 68.72, 41.15) -- ( 96.89, 41.15);

\path[draw=drawColor,line width= 0.4pt,line join=round,line cap=round] ( 93.76, 39.35) --
	( 96.89, 41.15) --
	( 93.76, 42.96);
\definecolor{drawColor}{RGB}{0,0,0}

\node[text=drawColor,anchor=base,inner sep=0pt, outer sep=0pt, scale=  0.90] at ( 82.80, 44.25) {$\dot x_n$};

\definecolor{drawColor}{RGB}{128,128,255}

\path[draw=drawColor,line width= 0.4pt,line join=round,line cap=round] (150.19, 41.15) -- (178.36, 41.15);

\path[draw=drawColor,line width= 0.4pt,line join=round,line cap=round] (175.23, 39.35) --
	(178.36, 41.15) --
	(175.23, 42.96);
\definecolor{drawColor}{RGB}{0,0,0}

\node[text=drawColor,anchor=base,inner sep=0pt, outer sep=0pt, scale=  0.90] at (164.27, 44.25) {$\dot x_{n+1}$};

\node[text=drawColor,anchor=base,inner sep=0pt, outer sep=0pt, scale=  0.90] at (123.54, 29) {$x_{n+1}-x_n$};

\node[text=drawColor,anchor=base,inner sep=0pt, outer sep=0pt, scale=  0.90] at ( 36.69, 13.45) {Ring of length $L$};

\node[text=drawColor,anchor=base,inner sep=0pt, outer sep=0pt, scale=  0.90] at (217.43, 13.45) {$N$ agents};

\node[text=drawColor,anchor=base,inner sep=0pt, outer sep=0pt, scale=  0.80] at ( 82.80, 7) {Agent $n$};

\node[text=drawColor,anchor=base,inner sep=0pt, outer sep=0pt, scale=  0.80] at (164.27, 7) {Agent $n+1$};

\end{scope}
\end{tikzpicture}
\end{center}\vspace{-2mm}
\caption{Scheme of the system with periodic boundary conditions. $x_n$ is the curvilinear position while $\Delta x_n=x_{n+1}-x_n$ is the spacing and $\dot x_n$ the speed of the particle with number $n$.}\label{fig1}
\end{figure}

We consider $N$ particles on a system of length $L$ with periodic boundary conditions. 
We denote in the following as $(x_n(t))_{n=1,\ldots,N}\in \mathbb R^N$ the cumulative curvilinear positions of the particles $n=1,\ldots,N$ at time $t\ge0$ (see Fig.~\ref{fig1}) and suppose that the particles are initially ordered by their index, i.e.
$$
x_1(0)\le x_2(0)\le...\le x_N(0)\le L+x_1(0).
$$

In the following model, the speed of a particle is a deterministic equilibrium speed function $V:s\mapsto V(s)$ depending on the spacing $s$ coupled to an additive stochastic noise. 
The speed function is related to as \emph{optimal velocity} in the traffic literature \cite{Bando1995},
We consider in the rest of the paper congested traffic states and the affine optimal velocity function 
\[
V(s)=\lambda(s-\ell),
\]
with $\lambda>0$ the inverse of the equilibrium time gap between the particles and $\ell\ge0$ their length. 
The time evolution of the particle with number $n = 1,\ldots,N$ is supposed to follow the stochastic ordinary differential equation
\be \label{model}
 \dot x_n(t) = \lambda (\Delta x_n(t)-\ell)+\xi_n(t), \qquad t \geq 0,
\ee
where $(\xi_n(t))_{t \geq 0}$ denotes the noise, $\dot x_n(t)$ denotes the tangential velocity,  
and the spacing between the particles are
\begin{equation}\label{def_spacing}
\left\{\begin{array}{lcll}\Delta x_n(t)&=& x_{n+1}(t)-x_n(t),&n=1,\ldots,N-1,\\[1mm]
\Delta x_N(t)&=&L+x_1(t)-x_N(t).\end{array}\right.
\end{equation}
Due to the system periodicity, the spacing sum $\sum_{n=1}^N\Delta_n(t)=L$ is conserved for all $t\ge0$. 
We could expect $L\ge N\ell$ to obtain a positive average speed of the particles. 
Such a condition is however mathematically not necessary to be well-defined.
We suppose that the noise is given by independent Ornstein-Uhlenbeck processes, i.e. 
\be \label{OU differential equation}
 \dd \xi_n(t) = -\beta\xi_n(t)\dd t + \sigma\dd W_n(t),
\ee
where $W_n(t)$, $n=1,\ldots,N$, are independent Wiener processes, $\beta > 0$ denotes the relaxation rate
and $\sigma \in \mathbb{R}$ the noise volatility, respectively.
Applying the It\^{o} formula to $C_n(t) = e^{\beta t} \xi_n(t)$ one finds that each $\xi_n(t)$
is given by
\be
\xi_n(t)=e^{-\beta t}\xi_n(0)+\sigma\int_0^t e^{\beta(s-t)}\dd W_n(s).
\label{OU}
\ee
Note that due to the noise introduced to initiate stop-and-go dynamics, the model does not ensure hard-core exclusion between the particles. 
Indeed, the noise being independent and unbounded, the probability that two particles overlap is not to exclude, especially at high density levels. 
More realistic features can be obtained by making the noise volatility proportional to the spacing \cite{Tordeux2016}.

Instead of \eqref{model}, 
we analyse the spacing difference of $x_n(t)$ to the space homogeneous solution $x_n^H(t)$, i.e.
\be\label{definition_Y}
y_n(t)=\Delta x_n(t)-\Delta x_n^{H}(t)
\ee
where the space homogeneous solution is the deterministic equilibrium configuration for which the vehicles are equispaced and have a constant speed at any time:
\be\label{solhomo}
 \begin{cases} x_n^H(t) = x_n^H(0) + t \lambda( L/N - \ell), \\ \Delta x_n^H(0) = L/N \end{cases}
\ee 
with $(x_n^H(t))_{n=1,\ldots,N}$ the cumulative curvilinear positions of a homogeneous system.
Representation (\ref{definition_Y}) has the advantage that it allows us to study the effects of noise around the equilibrium space homogeneous solution such as oscillating patterns and stop-and-go waves. 
We have for all $n=1,\ldots,N$
\[
 \dot y_n(t)=\lambda\big(y_{n+1}(t)-y_n(t)\big)+\xi_{n+1}(t)-\xi_n(t).
\]
This equation can be expressed by the system of stochastic ordinary differential equations
\be
\dot Y(t) = \lambda AY(t)+A\Xi(t),
\label{LS}
\ee 
where $Y(t)=\big[y_1(t),y_2(t),\ldots,y_N(t)\big]\tra\in\mathbb R^{N}$, $\Xi(t)=\big[\xi_1(t),\xi_2(t),\ldots,\xi_N(t)\big]\tra\in\mathbb R^{N}$ and  
$$A=\bmat -1&1&\\[-1.5mm]
&&\!\!\!\!\!\!\ddots\!\!\!\!\!\!\!\!\!\\[-4.5mm]
&&\!\!\!\!\!\!\!\!\!\ddots~\\[-4mm]
&&&~1\\1&&&-1\,\emat~~\in \text M_{N\times N}.$$ 
Let us stress that the processes $[x_1(t), \dots, x_n(t)]^{\top}$ obtained from \eqref{model} as well as $Y(t)$ obtained from \eqref{LS} both take values in $\R^N$, i.e.\ they are measured on an infinite lane using the cumulative arc length covered by each particle and by assuming, as given in (\ref{def_spacing}), that the spacing of the vehicle $N$ is $\Delta x_N(t)=L+x_1(t)-x_N(t)$.

\section{Solving the model}\label{solv}
Rewriting \eqref{LS} into the differential form
\be \label{Y differential form}
 \dd Y(t) = \big( \lambda A Y(t) + A \Xi(t) \big) \dd t
\ee 
shows that the noise $\Xi(t)$
enters in the definition of $Y(t)$ as an additional random drift parameter.
Hence $Y(t)$ cannot be a Markov process in its own. To overcome this difficulty we enlarge the state space from $\R^N$ to $\R^N \times \R^N$
by also taking the evolution of the noise $\Xi$ into account. 
In this way $Z := (Y,\Xi)$ becomes a Markov process with state space $\R^N \times \R^N$.

Indeed, using \eqref{OU differential equation}
combined with \eqref{Y differential form} we find that $Z(t) = (Y(t), \Xi(t))$ solves the system of stochastic differential equations
\be \label{SDE joint dynamics}
 \dd Z(t) = BZ(t) \dd t + G \dd W(t), \qquad Z(0) = (Y(0),\Xi(0)),
\ee
where $W(t) = (W_n(t))_{n = 1, \dots, 2N}$ is a family of independent Wiener processes and the $N \times N$ matrices $B,G$ are given by
\[
 B = \begin{pmatrix} \lambda A & A \\ 0 & - \beta 1_N \end{pmatrix},
 \qquad 
 G = \begin{pmatrix} 0 & 0 \\ 0 & \sigma 1_N \end{pmatrix},
\] 
where $1_N$ denotes the identity matrix acting on $\R^N$.

The particular form of \eqref{SDE joint dynamics}
shows that $Z$ is a $2N$-dimensional Ornstein-Uhlenbeck process and hence is given by
\be \label{OU joint dynamics solution}
 Z(t) = e^{tB} Z(0) + \int_0^t e^{(t-s)B}G \dd W(s).
\ee 
Following the general theory of Ornstein-Uhlenbeck processes 
(see, e.g., \cite{SY84, A15})
we find that $Z$ is a Feller process. 
Moreover, it is a Gaussian process whose 
characteristic function is, for $z,p \in \R^{2N}$, given by
\begin{align}
 \notag \mathbb{E}[ e^{i \langle p, Z(t) \rangle} \ | \ Z(0) = z ]
 &= \exp\left( i \langle z, e^{tB^{\top}}p \rangle - \frac{1}{2} \int_0^t \langle e^{s B^{\top}}p, G G^{\top} e^{s B^{\top}}p \rangle \dd s \right)
 \\ &= \exp\left( i \langle \mu_z(t), p \rangle - \frac{1}{2} \langle p , \Sigma(t) p \rangle \right) \label{characteristic function}
\end{align}
where its expectation $\mu_z(t)$ and covariance operator $\Sigma(t)$ are given by
\[ 
 \mu_z(t) = e^{t B}z, \qquad \Sigma(t) = \int_0^t e^{s B} G G^{\top} e^{s B^{\top}}\dd s.
\]
More generally one can also compute its covariance structure at different times.
\begin{lemma} \label{lemma1}
 For $t,s \geq 0$ it holds
 \[
  \mathrm{cov}(Z(t), Z(s)) = e^{tB}\int_0^{\min\{t,s\}} e^{- u B} G G^{\top} e^{- u B^{\top}} \dd u e^{s B^{\top}}.
 \]
\end{lemma}
The proof of Lemma \ref{lemma1} belongs to the classical literature of Ornstein-Uhlenbeck processes. See \cite{Revuz2013} for a general review.
As $Z$ is a Gaussian process, it is completely characterized by its expectation and covariance structure.
Based on the formulas of this section we can express all desired (statistical) quantities
in terms of the characteristic function and hence its mean and covariance structure.

\section{Stability analysis} \label{stab}
In this section we investigate the long-time behaviour of the mean
$\mathbb{E}[Z(t)]$, the limiting distribution of $Z(\infty)$, and finally invariant measures for the Markovian dynamics.
The results show that the process converges to a unique invariant measure which is on average a space homogeneous solution. 
However at the second order, the structure of the correlation functions and the presence of oscillating patterns allow to explain the presence of traffic waves. 
Such analysis crucially relies on the spectra of $A$ and $B$ which are, therefore, investigated first. 

\begin{proposition} \label{ergodicity}
 The matrix $A$ is diagonalizable with 
 eigenvalues 
 \[
  \omega_k = \gamma_k - 1, \qquad \gamma_k=e^{2\pi i\frac kN}, \qquad k = 0, \dots, N-1,
 \]
 and corresponding eigenvectors 
\be \label{eigenvectors A}
\mathbf{u}_k= \bmat \gamma_k^0 & \gamma_k^1 &\ldots &\gamma_k^{N-1}\emat^{\top}, \qquad k = 0,\dots N-1.
\ee
 The coefficients of the matrix exponential $e^{At}$ are given by
\begin{align}\label{formula matrix exponential A}
e^{At}(n,m)=\frac1N\sum_{k=0}^{N-1}\gamma_k^{n-m}e^{\omega_kt}, \qquad 1 \le n,m \le N
\end{align}
and it holds for each $y \in \R^{N}$
\begin{align}\label{ergodicity A}
 \left \| e^{At}y - \left(\frac{1}{N}\sum_{k=1}^{N}y_k \right) \bmat 1 \\ \vdots \\ 1 \emat \right\|_N \leq \sqrt{N}\|y\|_N e^{- 2 \sin\left( \frac{\pi}{N}\right)^2 t}, \qquad t \geq 0,
\end{align}
where $\| y \|_N^2 = \sum_{n=1}^{N}|y_n|^2$ denotes the euclidean norm on $\R^N$.
\end{proposition}
The proof of proposition~\ref{ergodicity} is a consequence of the circulant property of the matrix $A$, see \cite{Gray2006} for details.
Note that the coefficients of the exponential matrix $e^{At}$ in Eq.~(\ref{formula matrix exponential A}) are real-valued, even if expressed in the complex plane. 
Indeed, the imaginary parts vanish through the sum due to the oddness of the sine function.  
The complex parts come from the diagonalisation of $A$.
Yet the solution can be expressed in the real plan as well.
\begin{lemma}
The coefficients of the exponential of the matrix $A$ in the real plan are
\begin{equation}\label{formula matrix exponential A real plan}
e^{At}(n,m)=e^{-t}\sum_{l=0}^\infty \frac{t^{k(n,m,N)+lN}}{\big(k(n,m,N)+lN\big )!},\quad\text{with ~$k(n,m,N)=n-m\text{~mod~}N$}.
\end{equation}
for all $1\le n,m\le N$.
\end{lemma}
\begin{proof}
We can write $A=-I+D$, $D$ being a sparse matrix with an upper diagonal of ones (including the coefficient bottom left). 
The matrix $D^k$ is simply a shift of the diagonal $k$ step(s) to the left.
Then, remarking that $\sum_{k=1}^N D^{k+lN}$ is a matrix with one everywhere for all $l\in\mathbb N$ and using $e^A=e^{-1}e^D=e^{-1}\sum_k D^k/k!$ we obtain the expression above.
\end{proof}
The above expression  Eq.~(\ref{formula matrix exponential A real plan}) in the real plan is an infinite sum while the expression in the complex plane Eq.~(\ref{formula matrix exponential A}) solely requires finite computations. 
For numerical purpose, we prefer in the following using the finite sum Eq.~(\ref{formula matrix exponential A}) even if it implies using artificially complex numbers.

Next we continue with the analysis of the spectrum for $B$.
\begin{proposition}
 The matrix $B$ has eigenvalues 
 \begin{align}\label{eigenvalues B}
  ( \lambda \omega_{0}, \dots, \lambda \omega_{N-1}, - \beta , \dots, - \beta )
 \end{align}
 and corresponding eigenvectors
 \begin{align}\label{eigenvectors B}
  \left( \bmat \mathbf{u}_0 \\ 0 \emat, \dots \bmat \mathbf{u}_{N-1} \\ 0 \emat, \bmat - (\beta 1_N + \lambda A)^{-1}A e_1 \\ e_1 \emat, \dots,  \bmat - (\beta 1_N + \lambda A)^{-1}A e_N \\ e_N \emat \right),
 \end{align}
 where $e_1,\dots, e_N \in \R^N$ denote the canonical basis vectors in $\R^N$.
 In particular $B$ is diagonalisable and for each $z \in \R^{2N}$
 \begin{align}\label{ergodicity B}
  \left\| e^{Bt}z - \left( \frac{1}{N}\sum_{n=1}^{N}z_n \right) \bmat \mathbf{u}_0 \\ 0 \emat \right \|_{2N} \leq \sqrt{2N} \| z \|_{2N} e^{- \delta t}, \qquad t \geq 0,
 \end{align}
 where $\delta = \min\{ \beta, 2 \sin\left( \pi / N \right)^2 \} > 0$
 and $\mathbf{u}_0 = \bmat 1 & \hdots & 1 \emat^{\top} \in \R^N$.
\end{proposition}
\begin{proof}
The characteristic equation for $B$ is
\begin{align*}
 0 = \mathrm{det}\left( \bmat w 1_N & 0 \\ 0 & w 1_N \emat - \bmat \lambda A & A \\ 0 & - \beta 1_N \emat  \right)
 = \mathrm{det}( w1_N - \lambda A) \mathrm{det}(w1_N + \beta 1_N),
\end{align*}
whose solutions in $w \in \mathbb{C}$ are exactly \eqref{eigenvalues B}.
Let $\bmat y & \xi \emat^{\top} \in \R^{2N}$ be an eigenvector for the eigenvalue $\lambda \omega_k$, then
\[
 \lambda \omega_k \bmat y \\ \xi \emat = \bmat \lambda A & A \\ 0 & - \beta 1_N \emat \bmat y \\ \xi \emat = \bmat \lambda A y + A \xi \\ - \beta \xi \emat.
\]
Hence $\xi = 0$ and $y = \mathbf{u}_k$.
Similarly, let $\bmat y & \xi \emat^{\top} \in \R^{2N}$ be an eigenvector for the eigenvalue $-\beta$, then
\[
 - \beta \bmat y \\ \xi \emat = \bmat \lambda A y + A \xi \\ - \beta \xi \emat.
\]
Hence $\xi$ is arbitrary while $y$ satisfies
$(\beta + \lambda A)y = - A \xi$. 
Choosing $\xi\in\{e_1,\ldots,e_N\}$ shows that the eigenvectors are given by \eqref{eigenvectors B} and that
the corresponding eigenspaces span $\R^{2N}$, i.e. $B$ is diagonalisable. 
Concerning assertion \eqref{ergodicity B} we proceed similarly to \eqref{ergodicity A}. Let $\mathbf{v}_1, \dots, \mathbf{v}_{2N}$ be an orthonormal basis of eigenvectors of $B$
with $\mathbf{v}_1 = N^{-1/2}\bmat \mathbf{u}_0 & 0 \emat^{\top}$, and denote by $\varrho_1,\dots \varrho_{2N}$ the corresponding eigenvalues with 
$\varrho_n = \lambda \omega_{n-1}$, $n = 1, \dots, N$,
while $\varrho_n = - \beta$ for $n = N+1, \dots, 2N$.
For 
\[
z = \sum_{n=1}^{2N}\langle z, \mathbf{v}_n \rangle \mathbf{v}_n,
\]
we obtain
\[
e^{Bt}z = \sum_{n=1}^{2N}\langle z, \mathbf{v}_n \rangle e^{\varrho_n t}\mathbf{v}_n
\]
and hence 
\begin{align*}
    \left \| e^{Bt}z - \langle z, \mathbf{v}_1 \rangle \mathbf{v}_1 \right\|_{2N}
    &\leq \sum_{n=2}^{2N} | \langle z, \mathbf{v}_n \rangle| e^{\Re(\varrho_n) t}
    \\ &\leq e^{- \delta t} \sqrt{2N}\left( \sum_{n=2}^{2N} | \langle z, \mathbf{v}_n \rangle|^2 \right)^{1/2}
    \\ &\leq \sqrt{2N} \| z \|_{2N} e^{-\delta t},
\end{align*}
where we have used the Cauchy-Schwartz inequality and
\[
 \Re(\varrho_n) \leq - \delta, \qquad n = 2, \dots, 2N.
\]
Since $\langle z, v_1 \rangle \mathbf{v}_1 = \left( \frac{1}{N}\sum_{n=1}^{N}z_n \right) \bmat \mathbf{u}_0 & 0 \emat^{\top}$, the assertion is proved.
\end{proof}
Next we study the asymptotic behaviour of $Z(t)$ as $t \to \infty$.
\begin{theorem}
 It holds $Z(t) \fle_{t\fle\infty} Z(\infty)$ in law, 
 where $Z(\infty)$ is a Gaussian random variable on $\R^{2N}$ 
 with mean zero and covariance matrix
 \begin{align*}
     \Sigma(\infty) = \int_0^{\infty} e^{t B}G G^{\top} e^{tB^{\top}} \dd t.
 \end{align*}
\end{theorem}
\begin{proof}
 Using the characterization of convergence in law by characteristic functions (that is L\'evy's continuity Theorem, see e.g.\ \cite{Fristedt1996}), 
 it suffices to show that $\Sigma(\infty)$ is well-defined and that
 \begin{align}\label{convergence characteristic function}
  \lim \limits_{t \to \infty}\mathbb{E}[ e^{i \langle p, Z(t) \rangle} ] = \exp\left( - \frac{1}{2} \langle p, \Sigma(\infty)p \rangle \right), \qquad \forall p \in \mathbb{R}^{2N}.
 \end{align}
 Note that $\Sigma(\infty)$ is well-defined, if 
 \begin{align}\label{Sigma well defined}
  \int_0^{\infty} \left| \langle p, e^{Bt}G G^{\top} e^{B^{\top}t} q \rangle \right| \dd t < \infty, \qquad \forall p,q \in \mathbb{R}^{2N}.
 \end{align}
 Estimating first the scalar product and then the integral
 by Cauchy-Schwartz we arrive at
 \begin{align*}
  \int_0^{\infty} \left| \langle p, e^{Bt}G G^{\top} e^{B^{\top}t} q \rangle \right| \dd t
  &\leq \int_0^{\infty} \| G^{\top}e^{B^{\top}t} p\|_{2N} \| G^{\top}e^{B^{\top}t}q \|_{2N} \dd t 
  \\ &\leq \left( \int_0^{\infty} \| G^{\top}e^{B^{\top}t}p\|_{2N}^2 \dd t \right)^{1/2}\left( \int_0^{\infty} \| G^{\top}e^{B^{\top}t}q\|_{2N}^2 \dd t \right)^{1/2}.
 \end{align*}
 In order to show that these integrals are finite we first estimate $e^{Bt}G$
 in the Frobenius norm $\| \cdot \|_{\mathrm{F}}$ of a $2N \times 2N$
 matrix. Indeed, for each $p = \bmat p_1 & p_2 \emat^{\top} \in \R^{2N}$ we find 
 $Gp = \bmat 0 & \sigma p_2  \emat^{\top}$ and hence from \eqref{ergodicity B}
 applied to $z = Gp$
 \[
  \| e^{Bt}Gp \|_{2N} \leq \sqrt{2N} \| Gp \|_{2N} e^{- \delta t}
  \leq \sqrt{2N} \| G \|_{\mathrm{F}} \| p \|_{2N} e^{- \delta t},
 \]
 i.e. $\| e^{Bt}G \|_{\mathrm{F}} \leq \sqrt{2N} \| G \|_{\mathrm{F}}e^{- \delta t}$. From this we obtain 
 \[
  \| G^{\top}e^{B^{\top}t}p\|_{2N}
  \leq \| G^{\top} e^{B^{\top}t}\|_{\mathrm{F}} \| p \|_{2N}
  = \| e^{Bt}G \|_{\mathrm{F}} \| p \|_{2N}
  \leq \sqrt{2N} \| G \|_{\mathrm{F}}e^{- \delta t}\| p \|_{2N},
 \]
 which shows that \eqref{Sigma well defined} is satisfied.
 
 We proceed to prove \eqref{convergence characteristic function}.
 Using regular conditional distributions combined with \eqref{characteristic function} we find that
 \begin{align*}
     \mathbb{E}[ e^{i \langle p, Z(t) \rangle} ]
     &= \int_{\R^{2N}} \mathbb{E}[ e^{i \langle p, Z(t) \rangle} \ | \ Z(0) = z ] \mathbb{P}[ Z(0) \in \dd z]
     \\ &= e^{- \frac{1}{2} \langle p , \Sigma(t) p \rangle } \int_{\R^{2N}} e^{ i \langle e^{Bt}z, p \rangle } \mathbb{P}[ Z(0) \in \dd z].
 \end{align*}
 Using \eqref{Sigma well defined} we conclude that $\Sigma(t) \to \Sigma(\infty)$ as $t \to \infty$. Using \eqref{ergodicity B}
 we find 
 \[
  e^{Bt}z \longrightarrow \left( \frac{1}{N}\sum_{n=1}^{N} z_n \right) \left \langle p, \bmat \mathbf{u}_0 \\ 0 \emat \right \rangle
  = 0
 \]
 for $z \in Q = \{ w \in \R^{2N} \ | \ \sum_{n=1}^{N}w_n = 0 \}$.
 Then observing that 
\begin{align*}
\sum_{n=1}^{N}Z_n(0) 
&= \sum_{n=1}^Ny_n(0)
\\ &= \sum_{n=1}^N\big(\Delta x_n(0)-\Delta x^H_n(0)\big)
 \\ &= L+x_1(0)-x_N(0)+\sum_{n=1}^{N-1}\big(x_{n+1}(0)-x_n(0)\big)-\sum_{n=1}^N\Delta x^H_n(0)
 \\ &= L-L=0
\end{align*}
we find that $Z(0)$ belongs to $Q$ a.s.\ and hence
\begin{align*}
 \int_{\R^{2N}} e^{ i \langle e^{Bt}z, p \rangle } \mathbb{P}[ Z(0) \in \dd z]
 = \int_{Q}e^{ i \langle e^{Bt}z, p \rangle } \mathbb{P}[ Z(0) \in \dd z]
 \longrightarrow 1, \qquad t\to \infty.
\end{align*}
This proves \eqref{convergence characteristic function} and hence the assertion.
\end{proof}
This result shows that $\mathbb{E}[Z(t)] \longrightarrow 0$ as $t \to \infty$,
i.e. the whole dynamics tends asymptotically (in the mean) to the space homogeneous solution Eq.~(\ref{solhomo}). 
This means that the homogeneous solution is at the first order unconditionally stable for the stochastic model. 
This makes a clear difference with the classical deterministic approaches that describe stop-and-go waves by means of instability phenomena and phase transition \cite{Bando1995,Orosz2010}. 
In the stochastic approach, it is the structure of the correlation functions at the second order that allows explaining for the presence of traffic waves. 
Indeed, since $\Sigma(\infty) \neq 0$ the limiting law of $Z(\infty)$ is non-trivial and describes Gaussian fluctuations around the space homogeneous solution. 
Note that this law is also the unique invariant distribution for the process
(at least when restricted to the physically interesting configurations satisfying $\sum_{n=1}^{N}z_n = 0$).
As a consequence of previous result we find for the first component $Y$ 
\[
 \mathbb{E}[Y(t)] \longrightarrow 0
 \ \ \text{ and } \ \ Y(t) \xrightarrow{~d~} Y(\infty), \ \ \text{ as } \ \ t \to \infty,
\]
where $Y(\infty)$ is a Gaussian random variable $\R^{N}$ with covariance structure
\[
 \langle k, \Sigma_Y(\infty)p \rangle = \int_0^{\infty} \left\langle G^{\top}e^{B^{\top}s}\bmat k \\ 0 \emat, G^{\top}e^{B^{\top}s}\bmat p \\ 0 \emat \right \rangle \dd s.
\]
We close this section with a precise formula for $\mathbb{E}[Y(t)]$,
while the values for $\Sigma_Y(\infty)$ will be computed in the next section.
\begin{theorem}
 Let $Y$ be the solution of Eq.~(\ref{Y differential form}). One has
     \[
      \mathbb{E}[Y(t)] = e^{\lambda A t}\mathbb{E}[Y(0)]
      + (\beta 1_N + \lambda A)^{-1}\left( e^{-\beta 1_N t} - e^{\lambda A t} \right) A \mathbb{E}[\Xi(0)].
     \]
\end{theorem}
\begin{proof}
To simplify notation we let $\overline{Y}(t) = \mathbb{E}[Y(t)]$ and similarly $\overline{\Xi}(t) = \mathbb{E}[\Xi(t)]$.
Taking expectations in \eqref{LS} gives
\[
 \overline{Y}(t) = \lambda A \overline{Y}(t) - A \overline{\Xi}(t).
\]
Using \eqref{OU} so that $\overline{\Xi}(t) = e^{- \beta t} \overline{\Xi}(0)$ gives
\begin{align*}
 \overline{Y}(t) &= e^{\lambda A t}\overline{Y}(0)
 + \int_0^{t} e^{\lambda A(t-s)}e^{- \beta s}A\overline{\Xi}(s) \dd s
 \\ &= e^{\lambda A t}\overline{Y}(0) + e^{\lambda A t} \int_0^t e^{- (\beta + \lambda A)s} A\overline{\Xi}(0)\dd s
 \\ &= e^{\lambda A t} \overline{Y}(0) + e^{\lambda A t}(\beta 1_N + \lambda A)^{-1}( e^{- (\beta + \lambda A)t} - 1_{N}) A \overline{\Xi}(0)
  \\[2mm] &= e^{\lambda A t} \overline{Y}(0) + (\beta 1_N + \lambda A)^{-1}( e^{- \beta t}  - e^{\lambda A t}) A \overline{\Xi}(0),
\end{align*}
which proves the assertion. 
Note that here $\beta 1_N + \lambda A$ is invertible since $[\beta 1_N+\lambda A]X=0$ implies $X=(0,\ldots,0)\tra$ for all $\lambda,\beta>0$.
\end{proof}

\section{Covariance and autocovariance}\label{covN}
In the stochastic model, oscillation patterns in the correlation and autocorrelation of the particle spacing explain for the presence of collective stop-and-go waves in the system. 
The Gaussian framework of the model allows to obtain an explicit solution in stationary state for the correlation functions. 
Writing
\[
Y(t)=e^{\lambda At}C(t),
\]
with $C(t)$ a vector of size $N$, we obtain using Eq.~(\ref{LS})
$C'(t)=e^{-\lambda At}A\Xi(t)$.
One gets by integrating on $[0,t]$
\[ \textstyle
C(t)=C_0+\int_0^te^{-\lambda Au}A\Xi(u)\dd u.
\]
Here $C_0=C(0)=Y(0)$ and we obtain
\be
Y(t)=e^{\lambda At}C(t)=e^{\lambda At}Y(0)+\int_0^t e^{\lambda A(t-u)}A\Xi(u)\dd u,
\label{sol1}
\ee
or again, using the explicit solution $\xi_n(t)=e^{-\beta t}\xi_n(0)+\sigma\int_0^te^{\beta(u-t)}\dd W_n(u)$ for the Ornstein-Uhlenbeck processes,
\[
Y(t)=e^{\lambda At}Y(0)+R_0(t)+\sigma R(t),
\]
with
\[
 R_0(t)=\int_0^t e^{\lambda A(t-u)}Ae^{-\beta u}\dd u~\Xi(0),
\]
 and
\[
 R(t)=\int_0^t e^{\lambda A(t-u)}A\int_0^ue^{\beta(s-u)}\dd W(s)\dd u,
\]
$W(t)=(W_1(t),\ldots,W_N(t))\tra$ being a vector of independent Wiener processes.

We have
\[
\begin{array}{lcl}
R_0(t)&=&\int_0^t e^{-(\beta+\lambda A)u}\dd u~e^{\lambda At}A\Xi(0)\\[2mm]
&=&[\lambda A+\beta 1_N]^{-1}\big(1_N-e^{-\beta t}e^{-\lambda At}\big)e^{\lambda At}A\Xi(0)\\[2mm]
&=&[\lambda A+\beta 1_N]^{-1}\big(e^{\lambda At}A\Xi(0)-e^{-\beta t}\Xi(0)\big)~~\fle~~(0,\ldots,0)~~\text{as~~$t\fle\infty$},\end{array}
\]
since $e^{\lambda At}A$ and $e^{-\beta t}\Xi(0)$ tends to 0 as $t\fle\infty$, while $[\lambda A+\beta]X=0$ implies $X=(0,\ldots,0)\tra$ for all $\lambda,\beta>0$. 

We denote respectively in the following $\text{cov}_j(0)$ and $\text{cov}_0(\tau)$ the asymptotic covariance and autocovariance of the spacing difference of the particles
\[
\text{cov}(y_n(t),y_{n+j}(t))\fle_{t\fle\infty}\text{cov}_j(0),
\]
and
\[
\text{cov}(y_n(t),y_{n}(t+\tau))\fle_{t\fle\infty}\text{cov}_0(\tau).
\]
\begin{theorem}\label{defcov}
The asymptotic covariance of the spacing difference to the spacing difference of the particle $n+j$ ahead is for any particle $n=1,\ldots,N$,
\be
\text{\normalfont cov}_j(0)=\frac{\sigma^2}{2\beta N}\sum_{k=1}^{N-1}\frac{\gamma_k^{j}}{\lambda-\beta-\lambda\gamma_k}\left(\frac{(1-\gamma_k)^2}{\lambda-(\lambda+\beta)\gamma_k}-\frac{2\beta}{\lambda(\lambda+\beta-\lambda\gamma_k)}\right),
\label{cov}\ee
while the asymptotic autocovariance at time $\tau\ge0$ is 
\be
\text{\normalfont cov}_0(\tau)=\frac{\sigma^2}{2\beta N}\sum_{k=1}^{N-1}\frac{1}{\lambda-\beta-\lambda\gamma_k}\left(\frac{e^{-\beta \tau}(1-\gamma_k)^2}{\lambda-(\lambda+\beta)\gamma_k}-\frac{2\beta e^{-\lambda(1-\gamma_k)\tau}}{\lambda(\lambda+\beta-\lambda\gamma_k)}\right),
\label{ac}
\ee
with $\gamma_k=e^{2\pi i\frac kN}$ the $N$-roots of unity. 
\end{theorem} 
\begin{proof}
The autocovariance of the one-dimensional Ornstein-Uhlenbeck is
\be
\text{cov}(\xi_n(t),\xi_n(s))=\frac{\sigma^2}{2\beta}e^{-\beta(t+s)}\left(e^{2\beta\min\{t,s\}}-1\right),\ee
Using Eq.~(\ref{sol1}) by assuming $Y(0)=\Xi(0)=(0,\ldots,0)\tra$ in order to simplify the calculation and by remarking that $A+A\tra=-AA\tra$, the covariance of the process is\def\sl{\\[3mm]}
\be\begin{array}{l}		
\text{cov}(Y(t),Y(s))

=Ae^{\lambda At}\int_0^{t}\int_0^{s}e^{-\lambda Au}e^{-\lambda A\tra v}\text{cov}(\Xi(u),\Xi(v))\dd v\dd u~e^{\lambda A\tra s}A\tra\\[7mm]


=\sigma^2\big[\underbrace{Ae^{\lambda At}}_{\fle0}-\underbrace{Ae^{-\beta t}}_{\fle0}\big]\left[\beta 1_N+\lambda A\right]^{-1}\left[\lambda^2 \big(A\tra\big)^2-\beta^2 1_N\right]^{-1}\underbrace{e^{\lambda A\tra s}A\tra}_{\fle0}\\[7mm]
\displaystyle+\,\frac{\sigma^2}{\lambda}\big[\!\!\underbrace{e^{\lambda AA\tra t}}_{\fle (1/N)_{N^2}}\!\!-1_N\big]e^{\lambda A\tra (s-t)}\left[\lambda^2 \big(A\tra\big)^2-\beta^2 1_N\right]^{-1}\\[7mm]
\displaystyle-\,\frac{\sigma^2}{2\beta}\Big[\big[\underbrace{e^{-\beta s}Ae^{\lambda At}}_{\fle0}-e^{-\beta (s-t)}A\big]\left[\lambda A-\beta 1_N\right]^{-1}\\[7mm]
\displaystyle-\,\big[\underbrace{e^{-\beta s}Ae^{\lambda At}}_{\fle0}-\underbrace{e^{-\beta (t+s)}A}_{\fle0}\big]\left[\lambda A+\beta 1_N\right]^{-1}\Big]\left[\lambda A\tra+\beta 1_N\right]^{-1}A\tra.
\end{array}\label{eq:covarproof}\ee
The calculation details are provided in Appendix~1. 
We obtain asymptotically if $s=t+\tau$ with $\tau\ge0$,
\[
\begin{array}{lcl}
\displaystyle \lim_{t \to \infty}\text{cov}(Y(t),Y(t+\tau))&=&\displaystyle\frac{\sigma^2}{\lambda}\big[(1/N)_{N^2}-1_N\big]e^{\lambda A\tra \tau}\left[\lambda^2 \big(A\tra\big)^2-\beta^2 1_N\right]^{-1}\\[4mm]
&+&\displaystyle\frac{\sigma^2}{2\beta}e^{-\beta \tau}A\left[\lambda A-\beta 1_N\right]^{-1}\left[\lambda A\tra+\beta 1_N\right]^{-1}A\tra,
\end{array}
\]
with $(1/N)_{N^2}$ the $N\times N$ matrix with coefficients $1/N$ everywhere. Developing the matrix, one gets for any particle $n=1,\ldots,N$, the asymptotic covariance of the spacing difference to the spacing difference of the particle $n+j$ ahead
\[
\text{cov}_j(0)=\frac{\sigma^2}{2\beta N}\sum_{k=1}^{N-1}\frac{\gamma_k^{j}}{\lambda-\beta-\lambda\gamma_k}\left(\frac{(1-\gamma_k)^2}{\lambda-(\lambda+\beta)\gamma_k}-\frac{2\beta}{\lambda(\lambda+\beta-\lambda\gamma_k)}\right),
\]
while the asymptotic autocovariance at time $\tau\ge0$ is
\[
\text{cov}_0(\tau)=\frac{\sigma^2}{2\beta N}\sum_{k=1}^{N-1}\frac{1}{\lambda-\beta-\lambda\gamma_k}\left(\frac{e^{-\beta \tau}(1-\gamma_k)^2}{\lambda-(\lambda+\beta)\gamma_k}-\frac{2\beta e^{-\lambda(1-\gamma_k)\tau}}{\lambda(\lambda+\beta-\lambda\gamma_k)}\right),
\]
with $\gamma_k=e^{2\pi i\frac kN}$. 
\end{proof}
Note that the covariance and autocovariance  Eqs.~(\ref{cov}) and (\ref{ac}) are real-valued, even if expressed in the complex plane. 
Indeed, as for the exponential of the matrix $A$ Eq.~(\ref{formula matrix exponential A}), the imaginary parts vanish through the sum due to the oddness of the sine function. 
The merit of the complex expression, inherent to the diagonalisation of the matrix $A$, lies in obtaining exact numerical solutions. 
Explicit real-valued expressions are possible using series.

\begin{corollary}\label{defcor}
The correlation and autocorrelation 
\[ 
\text{\normalfont cor}_j(\tau)=\frac{\text{\normalfont cov}_j(\tau)}{\text{\normalfont cov}_0(0)} 
\]
do not depend on the parameter $\sigma$. 
\end{corollary}

\begin{figure}[!ht]
\begin{center}\small
\input{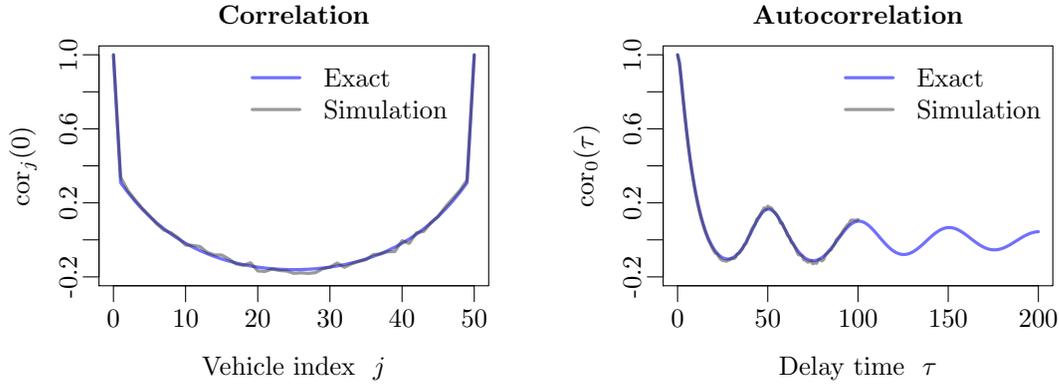}\vspace{-2mm}
\caption{\small Empirical and exact correlation and autocorrelation (see Eqs.~(\ref{cov}) and (\ref{ac})) for a system with $N=50$ particles in stationary state. $\lambda=1$ and $\beta=0.1$. 
The simulation results are computed using a Euler-Maruyama scheme with time step $\delta t = 0.01$~s. 
1e3 observations are measured after 1e5 units of simulation time.}
\label{fig2}
\end{center}
\end{figure}

The correlation with the neighbors and the autocorrelation in time of the spacing difference are presented Fig.~\ref{fig2} for $N=50$ particles, $\lambda=1$~s and $\beta=0.1$~s. 
Both theoretical solutions Eqs.~(\ref{cov}) and (\ref{ac}) and empirical value obtained by simulation are plotted.
The simulation results are computed using a Euler-Maruyama scheme with time step $\delta t = 0.01$~s. 
1e3 observations are averaged after 1e5 units of simulation time.
The correlation with the neighbors described a U-shape (see Fig.~\ref{fig2}, left panel). 
This is characteristic of propagation of a single wave in the system. 
In adequacy with the LWR theory and the Rankine--Hugoniot formula \cite{Richards1956,Lighthill1955}, the waves propagate backward in the system at the speed $v_w=-\lambda\ell$ while the particles travel in average at the speed $v = \lambda(L/N-\ell)$. 
Therefore, the wave period is $P=L/(v-v_w)=N/\lambda=50$~s  (see Fig.~\ref{fig2}, right panel).

\section{Covariance and autocovariance for the infinite system}
\label{cova}
In this section, we determine the covariance and autocovariance functions at the limit $N\rightarrow\infty$ of an infinite system. 
Such a limit allows to withdraw finite size effects and effects due to the periodic boundary conditions.

\begin{figure}[!ht]
\begin{center}\small
\input{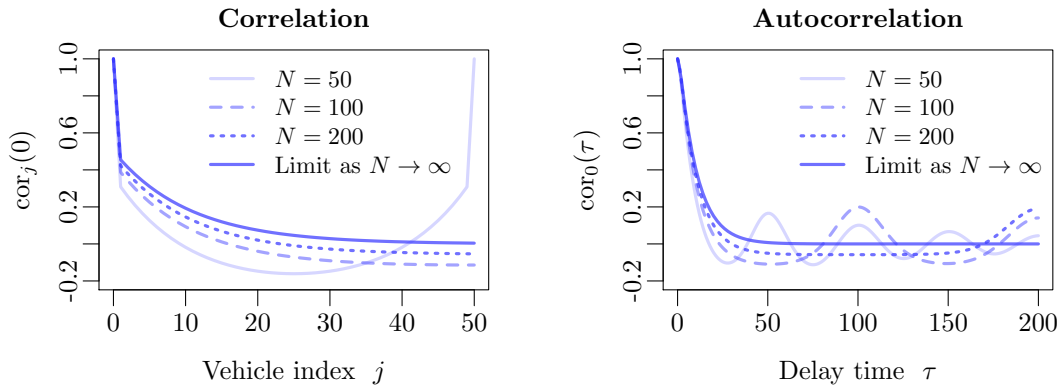}\vspace{-2mm}
\caption{\small Correlation and autocorrelation  in stationary state for systems with $N=50$, $100$, $200$ and at the limit $N\rightarrow\infty$ with $L/N$ constant (see Eqs.~(\ref{corasymp}) and (\ref{acasymp})) . $\lambda=1$ and $\beta=0.1$.}
\label{fig3}
\end{center}
\end{figure}

The covariance and autocovariance Eqs.~(\ref{cov}) and (\ref{ac}) at the limit $N\rightarrow\infty$ with $L/N$ constant are the Riemann integrals
\be
\text{cov}^\infty_j(\tau)=\frac{\sigma^2}{2\beta}\int_0^1F(e^{2\pi i t})\dd t=\frac{\sigma^2}{2\beta}\frac1{2\pi i}\int_{|z| = 1} \frac{F(z)}z\dd z
\label{covas}
\ee
with
\[
F(z)=\frac{1}{\lambda-\beta-\lambda z}\left(\frac{z^je^{-\beta \tau}(1-z)^2}{\lambda-(\lambda+\beta) z}-\frac{z^je^{\lambda(z-1)\tau}2\beta}{\lambda(\lambda+\beta-\lambda z)}\right).
\]
\begin{theorem}
The asymptotic correlation and autocorrelation of the spacing difference in stationary state are respectively at the limit $N\fle\infty$ with $L/N$ constant
\be
\mathrm{cor}^\infty_j(0)=\frac12\left(\frac\lambda{\lambda+\beta}\right)^j,\quad j>0,
\label{corasymp}
\ee
and
\be
\mathrm{cor}^\infty_0(\tau)=\frac{\lambda e^{-\beta\tau}-\beta e^{-\lambda\tau}}{\lambda-\beta},\quad\tau\ge0.
\label{acasymp}
\ee
\end{theorem}
\begin{proof}
We decompose the function $F(z)/z$ in simple elements to calculate the asymptotic autocovariance Eq.~(\ref{covas})
\[
\begin{array}{l}
\displaystyle\frac{F(z)}z=\frac{z^je^{-\beta\tau}}\lambda\left(\frac1{(\lambda-\beta)z}-\frac1{(\lambda-\beta)\left(z-\frac{\lambda-\beta}
\lambda\right)}+\frac1{(\lambda+\beta)\left(z-\frac\lambda{\lambda+\beta}\right)}\right)\\[5mm]
\displaystyle\hspace{2cm} -\; \frac{z^je^{\lambda(z-1)\tau}}\lambda\left(\frac{2\beta}{(\lambda^2-\beta^2)z}
-\frac1{(\lambda-\beta)\left(z-\frac{\lambda-\beta}\lambda\right)}
+\frac1{(\lambda+\beta)\left(z-\frac{\lambda+\beta}\lambda\right)} \right).
\end{array}
\]
Using the Cauchy formula 
\[
\frac1{2\pi i}\int_{|z| = 1} \frac{z^j}{z-\phi}\dd z = \begin{cases}\phi^j, & |\phi| < 1
\\ 0, & |\phi| > 1, \end{cases}
\] 
we obtain after calculations detailed in Appendix 2
\[
\text{cov}^\infty_j(0)=\left\{\begin{array}{ll}
\displaystyle\frac{\sigma^2}{\lambda\beta(\lambda+\beta)},&j=0,\\[4mm]
\displaystyle \frac{\sigma^2\lambda^{j-1}}{2\beta(\lambda+\beta)^{j+1}},&j>0,
\end{array}\right.
\]
Proceeding in the same way we find for the autocovariance Eq.~(\ref{ac}) at the limit $N,L\rightarrow\infty$
\[
\text{cov}^\infty_0(\tau)=\frac{\sigma^2}{\lambda\beta(\lambda^2-\beta^2)}\left(\lambda e^{-\beta\tau}-\beta e^{-\lambda\tau}\right).
\]
The asymptotic variance of the distance spacing is $\text{cov}^\infty_0(0)=\frac{\sigma^2}{\lambda\beta(\lambda+\beta)}$, while the asymptotic correlation and autocorrelation are respectively (see Fig.~\ref{fig3})
\[
\text{cor}^\infty_j(0)=\frac12\left(\frac\lambda{\lambda+\beta}\right)^j,\quad j>0,
\]
and
\[
\text{cor}^\infty_0(\tau)=\frac{\lambda e^{-\beta\tau}-\beta e^{-\lambda\tau}}{\lambda-\beta},\quad\tau\ge0.
\]
\end{proof}
The correlation in space and autocorrelation in time are both exponentially decreasing. 
The roles of the relaxation rate parameters $\lambda$ and $\beta$ in the correlation in space Eq.~(\ref{covas}) are antagonist while they are symmetric for the autocorrelation in time (see Eq.~(\ref{acasymp})).

In Fig.~(\ref{fig3}), the correlation and autocorrelation functions for the spacing difference in stationary state are plotted for $N=50$, $N=100$, $N=200$ and at the limit $N\fle\infty$ with $L/N$ constant for $\lambda=1$ and $\beta=0.1$. 
The correlation with the predecessors describes a U-shape due to the boundary condition, the correlation being one for $j=1$ and $j=N$ (Fig.~(\ref{fig3}), left panel). 
The correlation tends to increase in absolute value as the system size increases. 
The rescaled behaviors slightly differ according to $N$, tending to smooth U for small $N$ to step functions as $N$ increases.
The wave period of the autocorrelation in time is $P=N/\lambda=50$ for $N=50$, while it is $P=100$ and $P=200$ for $N=100$ and $N=200$ and is infinite at the limit $N\fle\infty$ (Fig.~(\ref{fig3}), right panel).

\section{Simulation results}\label{sim}
\begin{figure}[!ht]
\begin{center}\medskip
\input{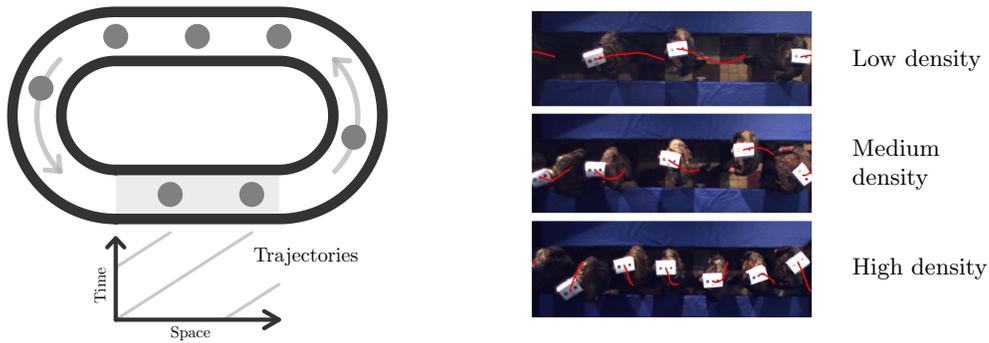}\smallskip
\caption{\small Schemes for the single-motion experiment and the collection of the trajectory data.}
\label{fig4}\vspace{5mm}
\end{center}
\end{figure}

\begin{figure}[!ht]
\begin{center}
\input{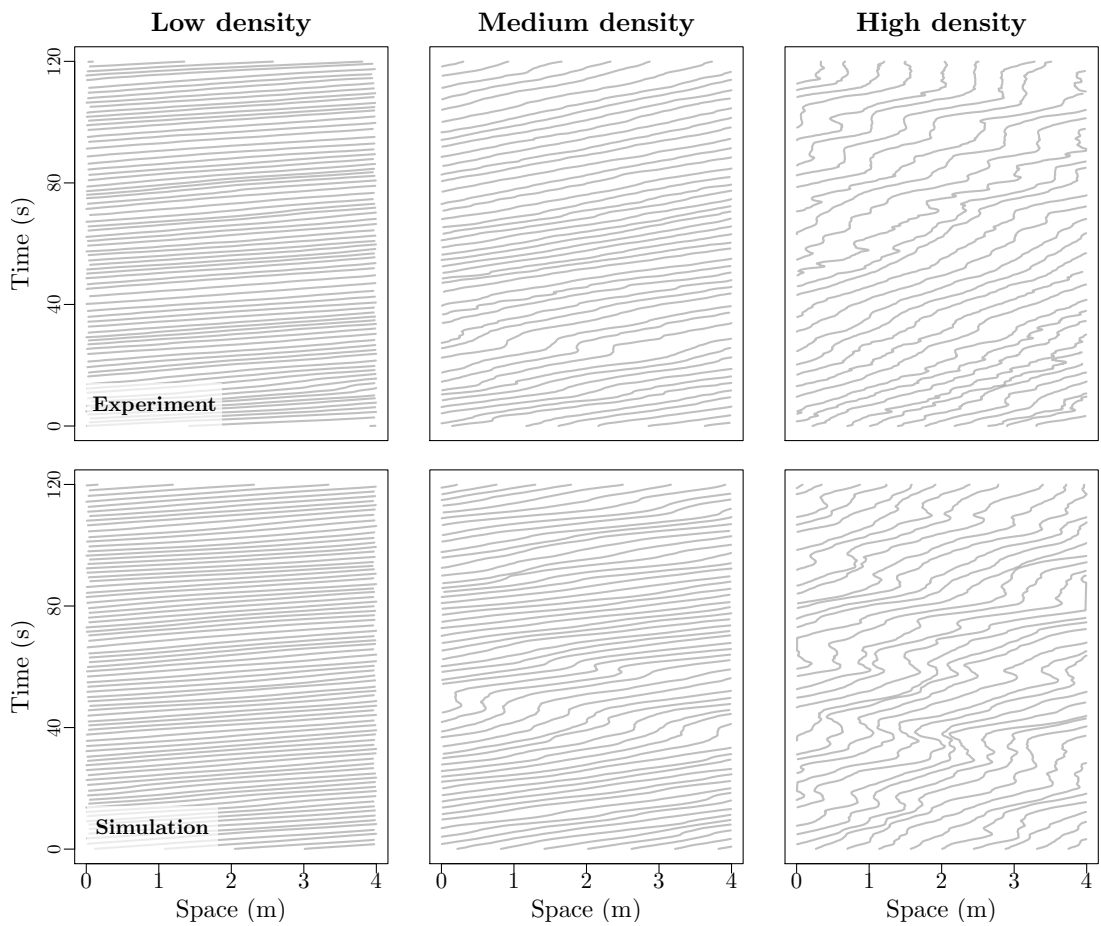}
\caption{\small Trajectories of single-file pedestrian motions with density levels 1~ped/m (left panels), 1.7~ped/m (central panels) and 2.3~ped/m (right panel). Top panels: Real experimental data. Bottom panels: Simulation of the calibrated stochastic pedestrian model. We observe stop-and-go waves for medium and high density levels in both real data and simulation.}
\label{fig5}\vspace{-7mm}
\end{center}
\end{figure}

Some simulation results are qualitatively compared to real single-file experiments obtained in laboratory conditions. The data come from experiments done on a quasi-circular geometry of length 27~m with soldiers in 2007 in Germany (see the schemes Fig.~\ref{fig4} and \cite{Portz2011,Tordeux2016} for details on the data). The stochastic pedestrian model is based on four parameters: the time gap inverse $\lambda$, the pedestrian length $\ell$, the noise relaxation rate $\beta$ and the noise volatility $\sigma$. The estimates of the parameters are $\tilde\lambda=0.98$~s$^{-1}$, $\tilde\ell=0.34$~m, $\tilde\beta=0.23$~s$^{-1}$ and $\tilde\sigma=0.09$~ms$^{3/2}$ \cite{Tordeux2016}. The trajectories for the experiments done with 28, 45 and 62 participants (corresponding to a density level of 1~ped/m, 1.7~ped/m and 2.3~ped/m) are plotted in Fig.~\ref{fig5}, top panels, while the simulated trajectories obtained with the stochastic model are shown  bottom panels. The simulation results are obtained using a Euler-Maruyama scheme with time step $\delta t=0.01$~s. The initial conditions are homogeneous. We rapidly observe spontaneous formation of stop-and-go waves for intermediate and high density levels in both experiments and simulations.

\section*{Appendix 1}\def\sl{\\[3mm]\dis}\def\slb{\\[7mm]\dis}\def\dis{\displaystyle}
The covariance of the spacing difference to the space homogeneous solution Eq.~(\ref{solhomo}) is, by using Eq.~(\ref{eq:covarproof}),
\be		
\begin{array}{c}\dis
\frac{2\beta}{\sigma^2}\left[Ae^{\lambda At}\right]^{-1}\text{cov}(\mathbf y(t),\mathbf y(s))\left[A\tra e^{\lambda A\tra s}\right]^{-1}\sl
=\int_0^{t}\int_0^{s}e^{-(\lambda A+\beta 1_N)u}e^{-(\lambda A\tra+\beta 1_N) v}\left(e^{2\beta\min\{u,v\}}-1\right)\dd v\dd u\slb
=\int_0^{t}\int_0^{u}e^{-(\lambda A+\beta 1_N)u}e^{-(\lambda A\tra+\beta 1_N) v}\left(e^{2\beta v}-1\right)\dd v\dd u\sl
+\int_0^{t}\int_u^{s}e^{-(\lambda A+\beta 1_N)u}e^{-(\lambda A\tra+\beta 1_N) v}\left(e^{2\beta u}-1\right)\dd v\dd u
\end{array}\label{A1}\tag{A1}
\ee

Finally, 
\be	
\begin{array}{l}\dis
\text{cov}(\mathbf y(t),\mathbf y(s))
=\frac{\sigma^2}{2\beta}Ae^{\lambda At}\int_0^{t}\left[e^{-(\lambda A+\beta 1_N)u}-e^{-\lambda (A+A\tra)u}\right]\dd u\sl
\left[\left[\lambda A\tra-\beta 1_N\right]^{-1}-\left[\lambda A\tra+\beta 1_N\right]^{-1}\right]~e^{\lambda A\tra s}A\tra\sl
-~\frac{\sigma^2}{2\beta}e^{-\beta s}Ae^{\lambda At}\int_0^{t}\left[e^{-(\lambda A-\beta 1_N)u}-e^{-(\lambda A+\beta 1_N) u}\right]\dd u\left[\lambda A\tra+\beta 1_N\right]^{-1}A\tra
\end{array}\label{A2}\tag{A2}
\ee
and we obtain Eq.~(\ref{eq:covarproof}) remarking that $A+A\tra=-AA\tra$ and $\left[\lambda A\tra-\beta 1_N\right]^{-1}-\left[\lambda A\tra+\beta 1_N\right]^{-1}
=2\beta\left[\lambda^2 \big(A\tra\big)^2-\beta^2 1_N\right]^{-1}$.

\section*{Appendix 2}
The covariance and autocovariance of the spacing difference at the limit $N\rightarrow\infty$ with $L/N$ constant are the Riemann integrals
\[
\text{cov}^\infty_j(\tau)=\frac{\sigma^2}{2\beta}\frac1{2\pi i}\int_{|z| = 1} \frac{F(z)}z\dd z,
\]
with
\[
\begin{array}{l}
\displaystyle\frac{F(z)}z=\frac{z^je^{-\beta\tau}}\lambda\left(\frac1{(\lambda-\beta)z}-\frac1{(\lambda-\beta)\left(z-\frac{\lambda-\beta}
\lambda\right)}+\frac1{(\lambda+\beta)\left(z-\frac\lambda{\lambda+\beta}\right)}\right)\\[5mm]
\displaystyle\hspace{2cm} -\; \frac{z^je^{\lambda(z-1)\tau}}\lambda\left(\frac{2\beta}{(\lambda^2-\beta^2)z}
-\frac1{(\lambda-\beta)\left(z-\frac{\lambda-\beta}\lambda\right)}
+\frac1{(\lambda+\beta)\left(z-\frac{\lambda+\beta}\lambda\right)} \right).
\end{array}
\]
In the following, the covariances and autocovariances are determined by using the Cauchy formula. 
We obtain the variance if $j=0$ and $\tau=0$
\be
\text{cov}^\infty_0(0)=\frac{\sigma^2}{2\lambda\beta}\left[
\frac1{\lambda-\beta}+\frac1{\lambda+\beta}-\frac{2\beta}{\lambda^2-\beta^2}\right]=\frac{\sigma^2}{\lambda\beta(\lambda+\beta)}.
\tag{A3}\ee
For $j>0$ and $\tau=0$, the covariance is
\be
\text{cov}^\infty_j(0)=\frac{\sigma^2}{2\lambda\beta}\frac{\left(\frac\lambda{\lambda+\beta}\right)^j}{\lambda+\beta}=\frac{\sigma^2\lambda^{j-1}}{2\beta(\lambda+\beta)^{j+1}},
\tag{A4}\ee
and $\text{cor}^\infty_j(0)=\frac12\big[\lambda/(\lambda+\beta)\big]^j$. 
For $j=0$ and $\tau\ge0$, the autocovariance is if $\left|\frac{\lambda-\beta}\lambda\right|\le1$, i.e.\ if $\beta\le2\lambda$,
\be
\begin{array}{lcl}
\displaystyle\text{cov}^\infty_0(\tau)&=&\displaystyle\frac{\sigma^2}{2\lambda\beta}\Bigg[
\frac{e^{-\beta\tau}}{\lambda+\beta}-\frac{2\beta e^{-\lambda\tau}}{\lambda^2-\beta^2}+\frac{e^{\lambda\big(\frac{\lambda-\beta}\lambda-1\big)\tau}}{\lambda-\beta}\Bigg]\\[5mm]
&=&\displaystyle\frac{\sigma^2}{2\lambda\beta}\Big[
\frac{e^{-\beta\tau}}{\lambda+\beta}-\frac{2\beta e^{-\lambda\tau}}{\lambda^2-\beta^2}+\frac{e^{-\beta\tau\tau}}{\lambda-\beta}\Big]~=~\displaystyle\frac{\sigma^2\big[\lambda e^{-\beta\tau}-\beta e^{-\lambda\tau}\big]}{\lambda\beta[\lambda^2-\beta^2]}.
\end{array}
\tag{A5}\ee
Similarly, we get if $\beta>2\lambda$
\[
\text{cov}^\infty_0(\tau)=\frac{\sigma^2}{2\lambda\beta}\Big[
\frac{e^{-\beta\tau}}{\lambda-\beta}+\frac{e^{-\beta\tau}}{\lambda+\beta}-\frac{2\beta e^{-\lambda\tau}}{\lambda^2-\beta^2}\Big]
=\frac{\sigma^2\big[\lambda e^{-\beta\tau}-\beta e^{-\lambda\tau}\big]}{\lambda\beta[\lambda^2-\beta^2]},
\]
and $\text{cor}^\infty_0(\tau)=[\lambda e^{-\beta\tau}-\beta e^{-\lambda\tau}]/[\lambda-\beta]$. 
Note that by taking $\lambda=\beta+\varepsilon$ and by calculating the autocovariance at the limit $\varepsilon\fle0$ we obtain 
\[\text{cov}^\infty_0(\tau)=\frac{\sigma^2e^{-\lambda\tau}}{2\lambda^3}[1+\lambda\tau]\] 
while $\text{cor}^\infty_0(\tau)=e^{-\lambda\tau}[1+\lambda\tau]$ if $\beta=\lambda$.

\end{document}